\documentclass[twocolumn,preprintnumbers]{revtex4}
\usepackage{amsmath}
\usepackage{epsf}
\usepackage{color}

\newcounter{nombre}

\begin{document}
\preprint{KUNS-2390}
\title{Breaking of $N=8$ magicity in $^{13}$Be}

\author{Yoshiko Kanada-En'yo}
\affiliation{Department of Physics, Kyoto University,
Kyoto 606-8502, Japan}

\begin{abstract}
Structure of $^{13}$Be was investigated with antisymmetrized molecular dynamics. 
The variation after spin and parity projections was performed. 
An unnatural parity $1/2^-$ state was suggested to be lower than $5/2^+$ state indicating that
vanishing of the $N=8$ magic number occurs in $^{13}$Be. 
A low-lying $3/2^+$ state with a $2\hbar\omega$ configuration was also suggested. 
Developed cluster structures were found in the intruder states. 
Lowering mechanism of the intruder states was discussed 
in terms of molecular orbitals around a $2\alpha$ core. 
\end{abstract}

\maketitle

\noindent
\section{Introduction}\label{sec:introduction}
One of the exotic phenomena discovered in unstable nuclei is
vanishing of neutron magic numbers. In neutron-rich Be isotopes, 
the breaking of $N=8$ magicity in $^{11}$Be has been known from the
abnormal spin and parity $1/2^+$ of the ground state.
The vanishing of the $N=8$ magic number in $^{12}$Be has been suggested by
slow $\beta$ decay to $^{12}$B \cite{Suzuki:1997zza} and it has been supported
by various experiments \cite{iwasaki00,iwasaki00b,Navin:2000zz,shimoura03,Pain:2005xw}.
The $N=8$ shell breaking has been suggested also 
in $^{11}$Li by experimental and theoretical works \cite{simon99}.
Even though $^{12}$Be and $^{11}$Li are neighboring nuclei, they have different 
characters and the shell breaking mechanism is not the same between two nuclei. 
One of the remarkable features of $^{12}$Be different from $^{11}$Li
is developed cluster structure, which 
plays an important role in the shell breaking in Be isotopes.

Cluster structures of Be isotopes have been intensively investigated 
in many theoretical works
\cite{OKABE,SEYA,OERTZEN,OERTZENa,Oertzen-rev,ITAGAKI,ENYObc,Dote:1997zz,AMDsupp-rev,ENYObe10,ENYObe11,ENYObe12,ARAI,OGAWA,Arai01,Arai:2004yf,Descouvemont01,Descouvemont02,Ito:2003px,Ito:2005yy,Ito:2008zza,Ito:2012zz}. 
Low-lying states in neutron-rich Be 
were described successfully with cluster models and molecular-orbital models
assuming two $\alpha$ clusters and surrounding neutrons. The formation of a 2$\alpha$ core in Be isotopes
was confirmed by the author and her collaborators 
with a method of antisymmetrized molecular dynamics which does not rely 
on {\it a priori} assumption of any clusters 
\cite{ENYObc,Dote:1997zz,AMDsupp-rev,ENYObe10,ENYObe11,ENYObe12}. These works revealed that 
$2\alpha$ structures are favored in neutron-rich Be isotopes where  
valence neutrons around the $2\alpha$ core play important roles.
To understand cluster features of low-lying states of neutron-rich Be,
a molecular orbital picture is helpful\cite{OKABE,SEYA,OERTZEN,OERTZENa,Oertzen-rev,ITAGAKI}. In the picture, 
molecular orbitals are formed by a linear combination of $p$
orbits around two $\alpha$ clusters, and valence neutrons occupy
the molecular orbitals.
A longitudinal positive-parity orbital is called ``$\sigma_{1/2}$-orbital". 
Since the $\sigma_{1/2}$-orbital has two nodes along the $\alpha$-$\alpha$ direction, 
it gains kinetic energy as the $2\alpha$ cluster develops. Note that
the $\sigma_{1/2}$-orbital corresponds to the ``$1/2[220]"$ orbit 
in the Nilsson model (a deformed shell model)\cite{Bohr-Mottelson}
and it originates in the $sd$-orbit in the spherical shell model limit.
In the molecular orbital picture,  
the ground states of $^{11}$Be and $^{12}$Be are regarded as the configurations 
with one and two neutrons in the $\sigma_{1/2}$-orbital, respectively, which correspond 
to the intruder configurations in terms of the spherical shell model. 
In other words, lowering mechanism of the intruder states in Be isotopes 
can be understood by the the energy gain of the $\sigma_{1/2}$-orbital 
in the developed 2$\alpha$ cluster structure.

It is then a challenging issue to investigate cluster features and shell evolution in
further neutron-rich Be isotopes near the drip line. 
Indeed, many experiments have been performed to observe energy spectra of $^{13}$Be 
(see Refs.~\cite{simon07,Kondo:2010zza} and
references therein).
Since $^{13}$Be is an unbound nucleus, all energy levels are resonance states. 
A resonance about 2 MeV above the neutron-decay threshold has been observed
in several experiments and it has been assigned to a $5/2^+$ state 
\cite{simon07,Kondo:2010zza,ostrowski92,korsheninnikov95,belozerov98,simon04}. 
However, other levels of $^{13}$Be have not been confirmed yet. 
The position of a $1/2^-$ state relative to the $5/2^+$ state is, in particular, a key problem concerning the breaking of $N=8$ magicity. 
Recently, energy spectra 
have been measured with invariant mass measurements of $^{12}$Be+$n$ by Simon {\it et al.} \cite{simon07}
and Kondo {\it et al.} \cite{Kondo:2010zza}. In the former work, they argued that 
an observed low-energy peak just above the $^{12}$Be+$n$ threshold is
described by $s$-wave virtual state contribution.
On the other hand, the latter work reported the existence of a $p$-wave resonance in the low-energy peak 
below the $5/2^+$ state and tentatively assigned it to a $1/2^-$ state
suggesting the breaking of $N=8$ magicity in $^{13}$Be.
However, a controversial claim was given \cite{Fortune:2010zz}, and the energy position of the $1/2^-$ state is still under
discussion.

If the inversion of single-particle levels does not occur in $^{13}$Be, the unnatural-parity
$1/2^-$ state should be higher than natural-parity $1/2^+$ and $5/2^+$ states. The question is whether or not
the inversion occurs and the $1/2^-$ state comes down to the low-energy region 
in $^{13}$Be. Provided that the inversion occurs, 
cluster structure may play an important role in the intruder $1/2^-$ state
as well as $^{11}$Be and $^{12}$Be.
Some theoretical calculations suggested possibility of a low-lying $1/2^-$ state in  $^{13}$Be\cite{labiche99,Kondo:2010zza}.
For instance, a shell model calculation for $^{13}$Be using the SFO interaction \cite{suzuki03}, 
which is adjusted to reproduce the 
parity inversion of $^{11}$Be, gives the $1/2^-$ ground state in $^{13}$Be \cite{Kondo:2010zza}. 
In microscopic cluster model calculations, there is only few
applications to $^{13}$Be  \cite{descouvemont95}
because of the large number of valence neutrons.


Our aim is to investigate structure of $^{13}$Be.
A method of energy variation after spin-parity projection (VAP) in the framework of
antisymmetrized molecular dynamics (AMD) \cite{ENYObc,AMDsupp-rev,ENYOa}
is applied to $^{13}$Be. The  AMD+VAP method has been already applied for studying 
$^{10}$Be\cite{ENYObe10}, $^{11}$Be\cite{ENYObe11}, and $^{12}$Be\cite{ENYObe12} 
and described successfully properties of ground and excited states.
In the energy levels of $^{13}$Be calculated by using the interaction that reproduces 
the parity inversion of $^{11}$Be, we will suggest 
a low-lying $1/2^-$ state below the $5/2^+$ state.
We show the coexistence of $0\hbar\omega$,  $1\hbar\omega$, and $2\hbar\omega$ states 
in a low-energy region of $^{13}$Be and suggest the breaking of $N=8$ magicity. 
We discuss cluster structures of $^{13}$Be and indicate that 
the calculated states of $^{13}$Be can be classified by molecular orbital configurations. 
The neutron shell breaking is discussed systematically in a chain of Be isotopes
focusing on cluster aspect.
 
This paper is organized as follows. 
In the next section, the formulation of the present calculation is explained. 
The results are shown in section \ref{sec:results} and 
discussions are given in section \ref{sec:discussions}. 
Finally, in section \ref{sec:summary}, a summary and an outlook are given.

\section{Formulation}\label{sec:formulation}
We describe $^{13}$Be with AMD wave functions by applying 
the VAP method. For the AMD+VAP method, the readers are referred to
Refs.~\cite{ENYObe10,ENYObe11,ENYObe12,ENYOe-c12}.
The method is basically the same as those applied to   
$^{10}$Be, $^{11}$Be, and $^{12}$Be. A difference in calculation procedures 
from Refs.~\cite{ENYObe10,ENYObe11,ENYObe12} is that we do not adopt
an artificial barrier potential in the present calculation 
which was used in the previous works.
To see effects of quasi-bound features of 
the last valence neutron on energy spectra,
we also adopt $^{12}$Be+$n$ wave functions by using the 
 $^{12}$Be core wave functions obtained with AMD+VAP for $^{12}$Be.

\subsection{AMD wave functions}

An AMD wave function is given by a Slater determinant of Gaussian wave packets;
\begin{equation}
 \Phi_{\rm AMD}({\bf Z}) = \frac{1}{\sqrt{A!}} {\cal{A}} \{
  \varphi_1,\varphi_2,...,\varphi_A \},
\end{equation}
where the $i$th single-particle wave function is written by a product of
spatial($\phi$), intrinsic spin($\chi$) and isospin($\tau$) 
wave functions as,
\begin{eqnarray}
 \varphi_i&=& \phi_{{\bf X}_i}\chi_i\tau_i,\\
 \phi_{{\bf X}_i}({\bf r}_j) & = &  \left(\frac{2\nu}{\pi}\right)^{4/3}
\exp\bigl\{-\nu({\bf r}_j-\frac{{\bf X}_i}{\sqrt{\nu}})^2\bigr\},
\label{eq:spatial}\\
 \chi_i &=& (\frac{1}{2}+\xi_i)\chi_{\uparrow}
 + (\frac{1}{2}-\xi_i)\chi_{\downarrow}.
\end{eqnarray}
$\phi_{{\bf X}_i}$ and $\chi_i$ are spatial and spin functions, and 
$\tau_i$ is the isospin
function fixed to be up (proton) or down (neutron). 
Accordingly, an AMD wave function
is expressed by a set of variational parameters, ${\bf Z}\equiv 
\{{\bf X}_1,{\bf X}_2,\cdots, {\bf X}_A,\xi_1,\xi_2,\cdots,\xi_A \}$.
The width parameter $\nu$ is chosen to be $\nu=0.17$ fm$^{-2}$  which is the same value as that used for 
$^{12}$Be in Ref.~\cite{ENYObe12}. 

\subsection{Variation after projection method}

Energy variation after spin and parity projections (VAP) in the AMD model space
is performed as is done in the previous studies of Be isotopes
\cite{ENYObe10,ENYObe11,ENYObe12}. 
For the lowest $J^\pi$ state,
the parameters ${\bf X}_i$ and $\xi_{i}$($i=1\sim A$) are varied to
minimize the energy expectation value of the Hamiltonian,
$\langle \Phi|H|\Phi\rangle/\langle \Phi|\Phi\rangle$,
with respect to the spin-parity eigen wave function projected 
from an AMD wave function; $\Phi=P^{J\pi}_{MK}\Phi_{\rm AMD}({\bf Z})$.
Here, $P^{J\pi}_{MK}$ is the spin-parity projection operator.
The energy variation is performed with a frictional cooling method \cite{AMDsupp-rev}.
Then the optimum AMD wave function
$\Phi_{\rm AMD}({\bf Z}^{J\pi})$,
which approximately describes the intrinsic wave function for 
a $J^\pi$ state, is obtained. 
For each $J^\pi=J_\alpha^{\pi_\alpha}$, the optimum parameters ${\bf Z}^{J_\alpha\pi_\alpha}$ are obtained.
After the VAP procedure, final wave functions are calculated by superposing 
the spin-parity eigen wave functions 
projected from all the AMD wave functions $\Phi_{\rm AMD}({\bf Z}^{J_\alpha\pi_\alpha})$
obtained by VAP for various $J_\alpha^{\pi_\alpha}$ states. 
Namely, the final wave functions for the $J^\pi$ states are expressed as, 
\begin{equation}\label{eq:diago}
|J^\pi\rangle=\sum_{\alpha,K} c^{J^\pi}(K,J_\alpha,\pi_\alpha) 
|P^{J\pi}_{MK}\Phi_{\rm AMD}({\bf Z}^{J_\alpha\pi_\alpha})\rangle,
\end{equation}
where the coefficients $c^{J^\pi}(K,J_\alpha,\pi_\alpha)$ are determined by
diagonalization of norm and Hamiltonian matrices.

\subsection{$^{12}$Be+$n$ model}
As shown later, the results obtained with AMD+VAP show that
low-lying states of $^{13}$Be can be interpreted as $^{12}$Be+$n$, where
the $^{12}$Be core is 
the intrinsic state of the 
$^{12}$Be($0^+_1$) having an intruder $2\hbar\omega$ configuration 
or that of $^{12}$Be($0^+_2$) with a normal $0\hbar\omega$
configuration. 
In reality, $^{13}$Be is an unbound nucleus and all states are resonances
above the $^{12}$Be+$n$ threshold.
In such a case, asymptotic behavior of the valence neutron wave function 
in the outer region can be important, in particular, for energy position of low angular-momentum states. 
However, the AMD method is not suitable to describe detailed behaviors of asymptotic regions, because 
a system is expressed by a Slater determinant of Gaussians 
and is treated in a bound state approximation in the AMD model. 
To see effects from spatial extension of the last neutron on energies of resonances,
we also calculate energy levels in a $^{12}$Be+$n$ model described below.

In the $^{12}$Be+$n$ model, 
we first apply AMD+VAP to $^{12}$Be to 
obtain intrinsic wave functions $\Phi(^{12}{\rm Be}:2\hbar\omega)$ of
$^{12}$Be($0^+_1$) and  $\Phi(^{12}{\rm Be}:0\hbar\omega)$ of $^{12}$Be($0^+_2$)
\cite{ENYObe12}.
Each intrinsic wave function $\Phi(^{12}{\rm Be}:\alpha)$ ($\alpha=0\hbar\omega, 2\hbar\omega$) 
is expressed by a AMD wave function and it is written by a Slater determinant. Then we add one 
neutron to the core wave functions   $\Phi(^{12}{\rm Be}:2\hbar\omega)$ and  $\Phi(^{12}{\rm Be}:0\hbar\omega)$. 
The additional neutron wave function in the $^{12}$Be+$n$ system is
described by a Gaussian wave packet located at a position ${\bf X}$ relative to the core.
A $^{13}$Be wave function for a $J^\pi$ state is described by a linear combination of 
$^{12}$Be+$n$ wave functions with various positions ${\bf X}$ as 
\begin{eqnarray}
\Psi^{J\pi} = &&\sum_{\alpha} \sum_{\sigma=\uparrow,\downarrow}\sum_{k} \sum_K\\
&& c^{J\pi}_{\alpha\sigma k K} P^{J\pi}_{MK}
{\cal A} \left\{ \Phi_{-\frac{{\bf X}_k}{13}}(^{12}{\rm Be}:\alpha) 
\psi_{n\sigma}(\frac{12}{13}{\bf X}_k)\right \},
\end{eqnarray}
where $\psi_{n\uparrow(\downarrow)}({\bf X})$ is a spin-up (spin-down)
neutron wave function with a Gaussian form with the width $\nu$ parameter localized at ${\bf X}$,
\begin{eqnarray}
 \psi_{n\uparrow(\downarrow)}({\bf X}) &=& \phi_{{\bf X}}\chi_{\uparrow(\downarrow)},\\
 \phi_{{\bf X}}&=&\left(\frac{2\nu}{\pi}\right)^{4/3}
\exp\bigl\{-\nu({\bf r}-\frac{{\bf X}}{\sqrt{\nu}})^2\bigr\}.
\end{eqnarray}
The $^{12}$Be wave function is shifted by $-{\bf X}_k/13$ to take into account
the recoil effect from the last neutron.  
For the total wave function, the antisymmetrization and the spin-parity projection
are performed as well as the superposition of basis wave functions.

In the present calculation, the intrinsic states $\Phi(^{12}{\rm Be}:\alpha) $
are deformed and their orientations are chosen to satisfy
$\langle x^2 \rangle \le \langle y^2 \rangle \le \langle z^2 \rangle$ and 
$\langle xy \rangle=\langle yz \rangle = \langle zx \rangle=0$. 
For the position ${\bf X}_k$ of the last neutron Gaussian wave function,
grid points in the $|x|\le 5$ fm and $|z|\le 5$ fm region on the $y=0$ plane 
are taken. The grid spacing is chosen to be 1 fm.

When an intrinsic wave function of the $^{12}$Be core is axial symmetric,  
${\bf X}_k$ on the $y$ plane is enough to take into account coupling of the last
neutron with all rotational band members of $^{12}$Be constructed from
the intrinsic wave function. Strictly speaking, 
the present $^{12}$Be wave functions are not axial symmetric, however,  
${\bf X}_k$ is restricted only on the $y=0$ plane to save numerical cost in the 
present calculation.
 
With the $^{12}$Be+$n$ model, we calculate $^{13}$Be energy spectra. 
Comparing the $^{12}$Be+$n$ model calculation with the AMD+VAP one, 
we will discuss, in particular, how the energy spectra can be modified from the AMD+VAP results
by improving the last neutron wave function.

\section{Results}\label{sec:results}

\subsection{Effective interaction}
We used the same effective nuclear interaction as that used
in Refs.~\cite{ENYObe11,ENYObe12}. It is the MV1 force \cite{MVOLKOV} 
 for the central force 
supplemented by a two-body spin-orbit force with the two-range Gaussian form 
same as that in the G3RS force \cite{LS}.
The Coulomb force is approximated using a seven-range
Gaussian form. 
We adopted the interaction parameters 
that are used  for $^{11}$Be and $^{12}$Be~\cite{ENYObe11,ENYObe12}. Namely, the 
Majorana, Bartlett, 
and Heisenberg parameters in the MV1 force 
are $m=0.65$, $b=0$, and $h=0$, respectively, and the 
spin-orbit strengths are taken to be $u_{I}=-u_{II}=3700$ MeV.
We denote this parameterization 
by the set (1) interaction.
To see the interaction dependence of the 
theoretical results, we also used the other parametrization (2) with
weaker spin-orbit forces, $u_{I}=-u_{II}=2500$ MeV. 

The energy levels of the excited states 
of $^{10}$Be, $^{11}$Be, and $^{12}$Be are
reproduced well by the AMD+VAP calculations with the set (1) interaction. 
Particularly, 
the breaking of the $N=8$ magicity
in $^{11}$Be and $^{12}$Be are successfully described with the set (1) interaction \cite{ENYObe11,ENYObe12}. 
In this paper, we mainly discuss the results calculated with the set (1). 

\subsection{Energy levels}

\begin{figure}[th]
\epsfxsize=7 cm
\centerline{\epsffile{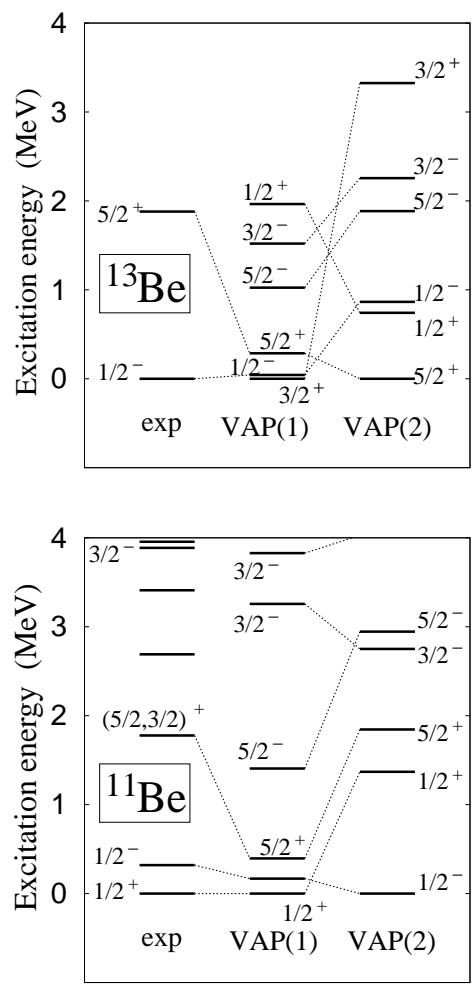}}
\caption{Energy levels of (upper)$^{13}$Be and (lower)$^{11}$Be.
Excitation energies are shown.
The calculated levels are obtained with AMD+VAP by using the set (1) and (2) interactions.
Results for $^{11}$Be obtained with AMD+VAP are taken from \cite{ENYObe11}.
The experimental data of $^{13}$Be are those in Ref.~\cite{Kondo:2010zza}.
\label{fig:be11-13spe}
}
\end{figure}

\begin{figure}[th]
\epsfxsize=7 cm
\centerline{\epsffile{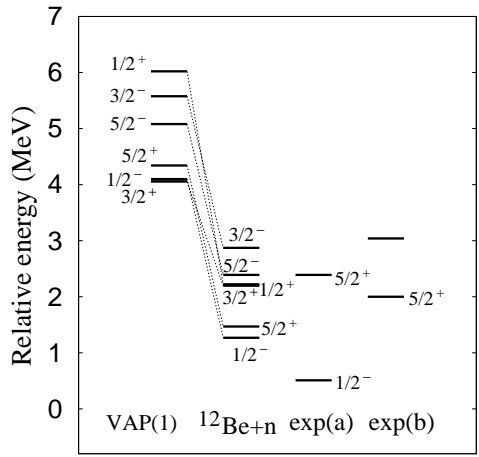}}
\caption{
Energy levels of $^{13}$Be calculated with AMD+VAP and those 
with the $^{12}$Be+$n$ model by using the set (1) interaction.
Energies relative to the $^{12}$Be-$n$ threshold are shown.
The calculated energies are measured from the theoretical 
$^{12}$Be-$n$ threshold energy,
 $-61.3$ MeV. Experimental energies are those taken from (a)
Ref.~\cite{Kondo:2010zza} and (b) Ref.~\cite{simon07}.
\label{fig:be13-Er}
}
\end{figure}

We applied AMD+VAP to $^{13}$Be and calculated 
$J^\pi$ states up to $J=5/2$.
The calculated energy levels of $^{13}$Be are shown in Fig.~\ref{fig:be11-13spe}
compared with experimental energy levels reported in Ref.~\cite{Kondo:2010zza}.
Energy levels of $^{11}$Be are also shown.
As mentioned in Ref.~\cite{ENYObe11}, the set (1) interaction reproduces 
the unnatural parity ground state $1/2^+$ in $^{11}$Be, while the set (2) interaction 
fails to describe the parity inversion.

In the result with the set (1) interaction, 
energy levels of $^{13}$Be are found to be out of the normal ordering. 
$1/2^-$ and $3/2^+$ states almost degenerate at the lowest energy and 
a $5/2^+$ state exists above them. 
As discussed later, the $1/2^-$ and $3/2^+$ states have 
dominantly $1\hbar\omega$ and $2\hbar\omega$ excited configurations, respectively,
while the  $5/2^+$ state is described by a normal $0\hbar\omega$ configuration.
The appearance of these intruder states in such the low-energy region 
suggests the breaking of $N=8$ magicity in $^{13}$Be 
as well as $^{11}$Be. 

In contrast, 
in the result with the set (2) interaction, the $5/2^+$ state is the lowest 
and the $1/2^-$ and $3/2^+$ states are higher than it as 
naively expected from the spherical shell model.

Although all states are resonances above the $^{12}$Be+$n$ threshold 
in a real $^{13}$Be system, 
they are treated in a bound state approximation
in the present AMD+VAP calculation. 
In principle, a resonance energy might decrease when the last neutron occupies 
a low angular-momentum orbit because of an extended neutron wave function in
an outer region. It means that asymptotic behavior of the valence neutron 
should be taken into account carefully for more detailed discussion of energy levels.
To see how the level ordering is affected by improving wave functions for the last neutron,
we calculated energy levels of $^{13}$Be also in the $^{12}$Be+$n$ model.

The $^{13}$Be energy spectra calculated with the $^{12}$Be+$n$ model by using the
set (1) interaction are shown in Fig.~\ref{fig:be13-Er}.
The calculated energies are measured from the theoretical value $-61.3$ MeV of
the $^{12}$Be-$n$ threshold energy, which is evaluated by diagonalizing
spin-parity eigen states projected from $\Phi(^{12}{\rm Be}:2\hbar\omega)$ and
$\Phi(^{12}{\rm Be}:0\hbar\omega)$.
 For all states, the $^{12}$Be+$n$ model calculation gives 
about  a few MeV lower energies than those obtained with AMD+VAP 
because of improving neutron wave functions.
The level ordering somehow changes from the VAP results, for instance, 
the $3/2^+$ state shifts up while the $1/2^+$ state comes down relatively.
However, 
the $^{12}$Be+$n$ calculation shows again the feature of the neutron magic number breaking 
that various spin and parity states degenerate 
in the low energy region.
In particular, it should be pointed that 
the $1/2^-$ state is the lowest consistently with the experimental
report by Kondo {\it et al.}\cite{Kondo:2010zza}.

In experimental and theoretical studies of $^{13}$Be, 
the $1/2^+$ state has been suggested to be a virtual state and contribute 
to the spectra near the $^{12}$Be+$n$ threshold energy  \cite{simon07,Kondo:2010zza,descouvemont95}. 
In the present calculation, model space is not enough to describe 
a virtual state. That may be the reason why the $1/2^+$ state is still 
higher than the $1/2^-$ and $5/2^+$ states even in the $^{12}$Be+$n$ model calculation.

\section{Discussions}\label{sec:discussions}
In the present AMD+VAP calculation, 
any clusters are not assumed in the model. Nevertheless, the results suggest that
cluster structures appear in $^{13}$Be as well as other Be isotopes. 
Moreover, the $^{13}$Be states obtained with AMD+VAP can be associated with 
$^{12}$Be+$n$ states, 
and their structures correspond to those 
of the $^{12}$Be+$n$ model calculation. 
Since each intrinsic wave function obtained 
by AMD+VAP is expressed with a single Slater determinant, AMD+VAP 
wave functions are useful to analyze intrinsic structures. 
We here discuss cluster features of $^{13}$Be by investigating 
AMD+VAP wave functions while focusing on  
the $2\alpha$ cluster and valence neutron configurations.
We also discuss systematics of low-lying states in neutron-rich Be isotopes 
in the molecular orbital picture.

\subsection{Intrinsic structures and clustering}

\begin{figure}[th]
\epsfxsize=7 cm
\centerline{\epsffile{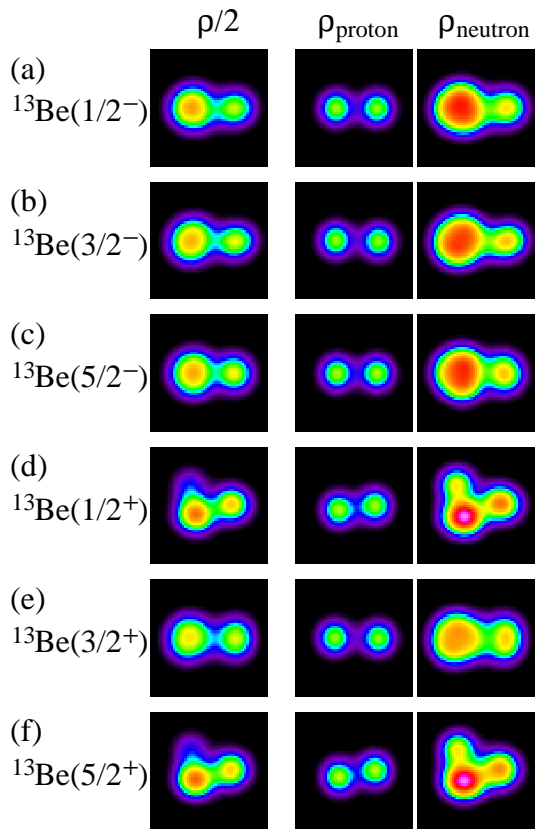}}
\caption{(Color on-line) Density distributions of the intrinsic states for $1/2^-$, $3/2^-$, 
$5/2^-$, $1/2^+$, $3/2^+$, and $5/2^+$  states in $^{13}$Be calculated with AMD+VAP
using the set (1) interaction.
The orientation of an intrinsic state is chosen so as to satisfy 
$\langle x^2 \rangle \le \langle y^2 \rangle \le \langle z^2 \rangle$ and 
$\langle xy \rangle=\langle yz \rangle = \langle zx \rangle=0$. 
The horizontal and vertical axes are set to the $z$ and $y$ axes, respectively.
Densities are integrated with respect to the $x$ axis.
Distributions of
matter, proton and neutron densities are shown left, middle,
and right, respectively.
\label{fig:be13dense}
}
\end{figure}

\begin{figure}[th]
\epsfxsize=7 cm
\centerline{\epsffile{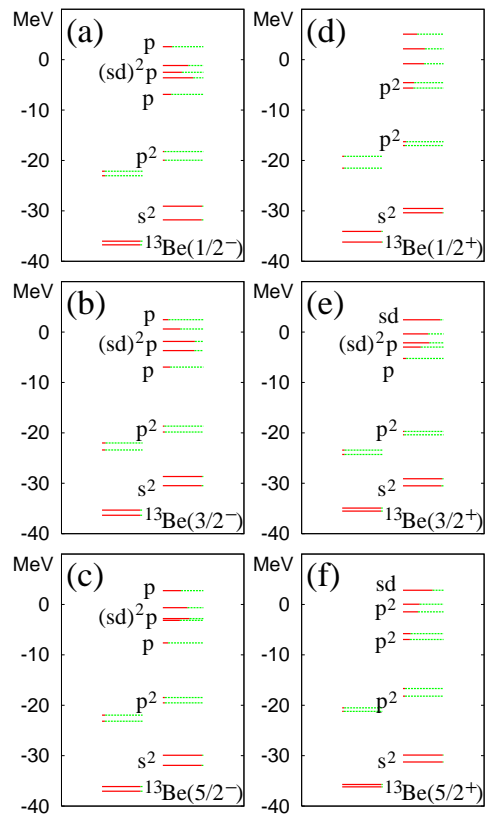}}
\caption{(Color on-line) Single-particle energies in the intrinsic 
wave functions  for  $^{13}$Be states,
 (a) $1/2^-$, (b) $3/2^-$, (c) $5/2^-$,  (d) $1/2^+$, (e) $3/2^+$, (f) $5/2^+$,   
calculated with AMD+VAP
using the set (1) interaction.
Fractions of positive- and negative-parity components are shown by red-solid and green-dotted lines, respectively. The labels, $s$, $p$, and $sd$ indicate association with shell-model orbits.
\label{fig:be13hfe}
}
\end{figure}

As seen in density distributions of intrinsic wave functions shown in Fig.~\ref{fig:be13dense},
a $2\alpha$ core is formed in $^{13}$Be. In particular, 
developed $2\alpha$ cluster structures with large deformations are found in  
negative-parity states, $^{13}$Be($1/2^-$), $^{13}$Be($3/2^-$), and $^{13}$Be($5/2^-$).
In positive-parity states, $^{13}$Be($1/2^+$) and $^{13}$Be($5/2^+$) show
relatively weaker cluster structures with smaller deformations while 
$^{13}$Be($3/2^+$) shows a remarkable $2\alpha$ structure similarly to the 
negative-parity states.

Since an intrinsic wave function for each state 
is given by a Slater determinant in the AMD+VAP calculation, 
we can analyze single-particle wave functions
and discuss neutron configurations as well as single-particle energies.
Single-particle wave functions and single-particle energies of an intrinsic state 
are calculated 
by diagonalizing the single-particle Hamiltonian matrix defined by analogy
to the Hartree-Fock theory as done in Refs.~\cite{Dote:1997zz,ENYObe10}.

As $^{13}$Be intrinsic states obtained with AMD+VAP are deformed and 
parity asymmetric, strictly speaking, spin and parity are not good quanta
in each single-particle wave function.
Nevertheless, single-particle wave functions can be 
associated with shell-model orbits from features of spatial distribution 
and ratios of positive- and negative-parity components.
Figure \ref{fig:be13hfe} shows single-particle energies in the intrinsic wave functions of $^{13}$Be states.
For each single-particle level, fractions of positive and negative-parity components are 
shown by a red-solid line and a green-dotted one, respectively. 
Labels indicate rough correspondence to shell-model orbits.
In the  $^{13}$Be($1/2^-$) state, the proton orbits and the lowest four neutron orbits form a $2\alpha$ core.
Among five valence neutrons around the $2\alpha$ core, three of them occupy $p$-like orbits
and the other two occupy $sd$-like orbits. 
Thus, the $^{13}$Be($1/2^-$) roughly corresponds to a $(p)^{-1}(sd)^2$ configuration 
on the neutron $p$-shell
and it is regarded as a $1\hbar\omega$ excited configuration. In a similar way, also 
the $^{13}$Be($3/2^-$) and $^{13}$Be($5/2^-$) states
have dominantly $(p)^{-1}(sd)^2$ configurations, 
and they are regarded as $1\hbar\omega$ members as well as $^{13}$Be($1/2^-$).
On the other hand, the $^{13}$Be($5/2^+$) contains mainly a $(sd)^1$ configuration and corresponds to 
the normal $0\hbar\omega$ configuration. 
In contrast to the $^{13}$Be($5/2^+$), the $^{13}$Be($3/2^+$) contains dominantly a $2\hbar\omega$ excited configuration of 
a $(p)^{-2}(sd)^3$ configuration. It is surprising that $0\hbar\omega$, $1\hbar\omega$, and $2\hbar\omega$ states
almost degenerate in the low-energy region in $^{13}$Be. 

For the $^{13}$Be($1/2^+$) obtained with AMD+VAP, we can not associate 
valence neutron wave functions with shell model orbits because parities are strongly mixing 
and spatial behaviors show no specific feature analogous to shell-model orbits.
Furthermore, this state may not correspond to the $s$-wave virtual state, which has been 
 suggested near the $^{12}$Be+$n$ threshold. 

\begin{figure}[th]
\epsfxsize=7 cm
\centerline{\epsffile{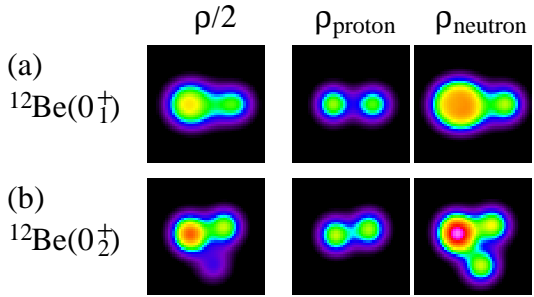}}
\caption{(Color on-line) Density distributions of the intrinsic states for 
the $0^+_1$ and $0^+_2$ states of $^{12}$Be calculated with AMD+VAP using the set (1) interaction.
Same as Fig.~\ref{fig:be13dense}.
\label{fig:be12dense}
}
\end{figure}

\begin{figure}[th]
\epsfxsize=7 cm
\centerline{\epsffile{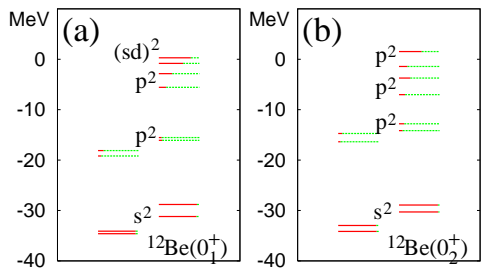}}
\caption{
(Color on-line) Single-particle energies in the intrinsic 
wave functions of $^{12}$Be states.
Same as Fig.~\ref{fig:be13hfe}.
\label{fig:be12hfe}
}
\end{figure}

Let us compare deformations and cluster structures in $^{13}$Be with those in $^{12}$Be.
Figures ~\ref{fig:be12dense} and ~\ref{fig:be12hfe} show density distributions 
and single-particle energies of the intrinsic wave functions $\Phi(^{12}{\rm Be};2\hbar\omega)$
for $^{12}$Be$(0^+_1)$  
and $\Phi(^{12}{\rm Be};0\hbar\omega)$
for $^{12}$Be$(0^+_2)$ obtained with AMD+VAP. 
As discussed in Ref.~\cite{ENYObe12}, $^{12}$Be$(0^+_1)$ is dominantly the 
$2\hbar\omega$ intruder state having a developed cluster structure with a large deformation,
while 
$^{12}$Be$(0^+_2)$ described dominantly by the normal $0\hbar\omega$ configuration
has a weaker cluster structure and a smaller deformation than those of $^{12}$Be$(0^+_1)$.
It is found that  
the large deformations in $^{13}$Be($1/2^-$), $^{13}$Be($3/2^-$), $^{13}$Be($5/2^-$), and $^{13}$Be($3/2^+$)
are  similar to that of $^{12}$Be$(0^+_1)$, while smaller deformations in 
$^{13}$Be($1/2^+$) and $^{13}$Be($5/2^+$) 
are associated with $^{12}$Be$(0^+_2)$.
Then,
the $^{13}$Be states can be interpreted as $^{12}$Be+$n$ states by considering 
an aditional neutron on the $^{12}$Be cores.
That is, $^{13}$Be($1/2^-$), $^{13}$Be($3/2^-$), and  $^{13}$Be($5/2^-$) 
have the $^{12}$Be($2\hbar\omega$) core with a neutron in a $p$-orbit.
$^{13}$Be($3/2^+$) is regarded as the $^{12}$Be($2\hbar\omega$) core and a $sd$-orbit neutron,
while $^{13}$Be($5/2^+$) is interpreted
as the $^{12}$Be($0\hbar\omega$) core and a $sd$-orbit neutron. 
Here coupling of the last neutron with the $^{12}$Be($2\hbar\omega$) core is not weak coupling 
but strong coupling where the neutron is moving around the largely deformed core.

The $^{12}$Be+$n$ features in the AMD+VAP results are supported also by the
 $^{12}$Be+$n$ model calculation.
Indeed, 
in the $^{12}$Be+$n$ model calculation, 
$^{13}$Be($1/2^-$), $^{13}$Be($3/2^-$), $^{13}$Be($5/2^-$), and $^{13}$Be($3/2^+$) 
contain mainly $^{12}$Be($2\hbar\omega$)+$n$ components, while $^{13}$Be($5/2^+$) and 
$^{13}$Be($1/2^+$) are approximately described by $^{12}$Be($0\hbar\omega$)+$n$ wave functions.

\subsection{Molecular orbital picture}

In theoretical works on cluster structures of Be isotopes
\cite{OKABE,SEYA,OERTZEN,OERTZENa,Oertzen-rev,ITAGAKI,ENYObc,Dote:1997zz,AMDsupp-rev,ENYObe10,ENYObe11,ENYObe12,ARAI,OGAWA,Arai01,Arai:2004yf,Descouvemont01,Descouvemont02,Ito:2003px,Ito:2005yy,Ito:2008zza,Ito:2012zz},
low-lying states of $^9$Be, $^{10}$Be, $^{11}$Be, and  $^{12}$Be
have been successfully described in terms of molecular orbitals around a
2$\alpha$ core. 
As mentioned, the present work suggests the $2\alpha$ core formation 
in $^{13}$Be as well as in other Be isotopes.
We here extend the molecular orbital description to $^{13}$Be and 
discuss structures of Be isotopes systematically. 

\begin{figure}[th]
\epsfxsize=7 cm
\centerline{\epsffile{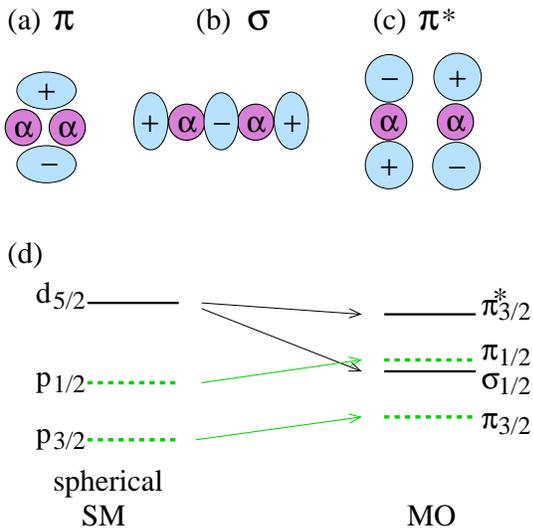}}
\caption{(Color on-line) (a)(b)(c) Schematic figures for molecular orbitals around a $2\alpha$ core.
(d) A schematic figure for evolution of single-particle level ordering 
from spherical shell-model levels to molecular orbital levels.
\label{fig:orbital}
}
\end{figure}

In the molecular orbital picture for a $2\alpha$ system, 
molecular orbitals are formed by a linear combination of $p$
orbits around two $\alpha$ clusters, and valence neutrons occupy
the molecular orbitals around the $2\alpha$ core 
\cite{OKABE,SEYA,OERTZEN,OERTZENa,Oertzen-rev,ITAGAKI}.
A negative-parity orbital constructed by $p$-orbits perpendicular to 
the $\alpha$-$\alpha$ direction is called a ``$\pi$-orbital", and
a positive-parity orbital from $p$-orbits parallel to 
the $\alpha$-$\alpha$ direction is called a ``$\sigma$-orbital" 
(Figs.~\ref{fig:orbital}(a) and .~\ref{fig:orbital}(b)). 
We call the other positive-party orbital given by $p$-orbits perpendicular to 
the $\alpha$-$\alpha$ direction a ``$\pi^*$-orbital" in analogy to electron orbitals in 
atomic molecular systems (Fig.~\ref{fig:orbital}(c)). 


To understand the breaking of magicity in a chain of Be isotopes, 
single-particle levels in the molecular orbital model (MO levels) is useful as discussed, for instance, 
in Ref.~\cite{Oertzen-rev}.
In the molecular orbital models \cite{SEYA,OERTZEN,OERTZENa,Oertzen-rev}, 
single-particle levels are evaluated as functions of the $\alpha$-$\alpha$ distance. They are 
smoothly connected from the one-center limit to the two-center limit, and those 
in the intermediate region correspond to molecular orbitals.
In addition to the spatial configurations ($\pi$, $\sigma$, and $\pi^*$),  
molecular orbitals are specified by the angular momentum $\Omega\equiv j_z$ 
projected on to the symmetric axis $z$.
$\Omega$ for the $\sigma$-orbital is $\Omega=1/2$, and we use the notation $\sigma_{1/2}$.
For $\pi$- and $\pi^*$-orbitals, $\Omega=1/2$ and $3/2$ are possible. 
Due to the spin-orbit force, $\pi$-orbitals split into the $ls$-favored $\pi_{3/2}$-orbital and
the $ls$-unfavored $\pi_{1/2}$-orbital, and  $\pi^*_{3/2}$-orbitals do into the $\pi^*_{3/2}$- 
and $\pi^*_{1/2}$-orbitals.
Note that the present notations,  $\pi_{3/2}$, $\pi_{1/2}$,  $\sigma_{1/2}$, and  $\pi^*_{3/2}$ correspond to 
the labels $\pi 3/2^-(g)$, $\sigma 1/2^-(g)$, $\sigma 1/2^+(u)$, and $\pi 3/2^+(u)$ in Fig.~15 of Ref.~\cite{Oertzen-rev}, and 
the labels $(3u,1)$, $(1u,2)$, $(1g,2)$, and $(1g,2)$ in Ref.~\cite{SEYA}, respectively.
In the spherical shell model limit, the $\pi_{3/2}$- and $\pi_{1/2}$-orbitals lead to the $p_{3/2}$- and $p_{1/2}$-orbits, respectively, while the $\sigma_{1/2}$- and $\pi^*_{3/2}$-orbitals go 
to the $d_{5/2}$-orbit. 
The MO levels correspond well to the single-particle levels in the two-center shell model \cite{TC-SM}.
Moreover, when a $\alpha$-$\alpha$ distance is not large, the MO levels 
are associated with the Nilsson levels of the deformed shell model\cite{Bohr-Mottelson}.

One of the important features of molecular orbitals is that the $\sigma$-orbital 
has two nodes along the $\alpha$-$\alpha$ direction and it 
gains the kinetic energy as the 2$\alpha$ cluster develops.
The breaking of $N=8$ magicity in $^{11}$Be and $^{12}$Be 
is understood by the energy gain of the $\sigma_{1/2}$-orbital 
in developed 2$\alpha$ systems.

Let us describe structures of $^{13}$Be with 
a $2\alpha$ core and five valence neutrons in terms of molecular orbitals.
In the developed cluster states of $^{13}$Be obtained with AMD+VAP, 
neutron wave functions 
are associated with molecular orbitals.
For instance, in the $1/2^-$ state, three neutron wave functions correspond to the $\pi$-orbital
and two neutron orbitals are associated with the $\sigma$-orbital. This state
is described by a molecular configuration, $\pi_{3/2}^2\pi_{1/2}^1\sigma_{1/2}^2$.
In a similar way, the $3/2^+$ state is understood by a $\pi_{3/2}^2\sigma_{1/2}^2\pi_{3/2}^*$ configuration.
In both cases, two neutrons occupy the $\sigma_{1/2}$-orbital. 
Then, the reason for the low-lying intruder states in $^{13}$Be can be understood again 
by the lowering $\sigma_{1/2}$-orbital in the developed cluster structures. 
In other words, $\sigma_{1/2}$-orbital neutrons play an important role for the breaking of 
$N=8$ magicity in $^{13}$Be as well as $^{12}$Be. 
As for the $5/2^+$ state in $^{13}$Be, since it has a weaker cluster structure and a smaller deformation, 
its neutron configurations should be associated with spherical shell-model orbits rather than 
molecular orbitals. Nevertheless, taking into account the correspondence 
of the $\pi_{3/2}$-, $\pi_{1/2}$-, and $\sigma_{1/2}$-orbitals to the $p_{3/2}$-, $p_{1/2}$-, and $d_{5/2}$-orbits, we here temporarily assign ``$\pi_{3/2}^2\pi_{1/2}^2\sigma_{1/2}^1$" to the $5/2^+$ state 
in the following discussion.

Energy spectra of Be isotopes can be understood 
systematically according to the ordering of MO levels. 
In the MO level ordering, the key feature is 
that such the orbitals as $\sigma$- and $\pi^*$-orbitals with nodal structures parallel to the $2\alpha$ direction gain 
kinetic energy in a developed 2$\alpha$ system. 
As mentioned, the $\sigma$-orbital with two nodes gains the energy as the cluster develops. 
Also the $\pi^*$-orbitals with one node gain some 
kinetic energy in a developed cluster system. Differently from the $\sigma$- and $\pi^*$-orbitals, 
the $\pi$-orbitals have no node and their energies increase relatively. 
Consequently, as the $2\alpha$ cluster develops, the ordering of single-particle levels changes 
from spherical shell model orbits 
as shown in Fig.~\ref{fig:orbital} (d).
In this scenario, the breaking of the neutron magicity in Be isotopes 
occurs due to the intruder $\sigma_{1/2}$-orbital which comes down
below the $\pi_{1/2}$-orbital in the developed $2\alpha$ systems.
As a result, the $N=8$ shell gap disappears and the level ordering based on a 
spherical shell model picture is no longer valid in neutron-rich Be.
Instead, the level ordering of molecular orbitals works rather well to understand 
energy spectra of Be isotopes. 
From the fact that $^{11}$Be$(1/2^+)$ and $^{11}$Be$(1/2^-)$ almost degenerate, 
one may expect, in the first-order approximation, that 
the $\pi_{1/2}$- and $\sigma_{1/2}$-orbitals almost degenerate in the MO levels.
In the new ordering of the MO levels on a $2\alpha$ core, the $\pi_{3/2}$-orbital should 
be the lowest,  $\sigma_{1/2}$- and $\pi_{1/2}$-orbitals compose the second group 
(called a $\sigma_{1/2}$-$\pi_{1/2}$ shell in the present paper), 
and the $\pi^*_{3/2}$-orbital may exist above them (Fig.~\ref{fig:orbital} (d)). 

Let us review the molecular orbital configurations of Be isotopes. 
In Table \ref{tab:beiso}, configurations for valence neutrons around a $2\alpha$ core 
for band-head states of Be isotopes are summarized.
For $^{10}$Be,  
$^{10}$Be($0^+_1$), $^{10}$Be($1^-$), and $^{10}$Be($0^+_2$) are described by the valence neutron configurations of  
$\pi_{3/2}^2$,  $\pi_{3/2}\sigma_{1/2}$, and $\sigma_{1/2}^2$ meaning two neutrons in the 
$\pi$-orbital, one neutron in the $\pi_{3/2}$-orbital and the other neutron in the $\sigma_{1/2}$-orbital, and 
two neutrons in the $\sigma_{1/2}$-orbital, respectively. 
In a similar way,  
$^{11}$Be($1/2^+$),$^{11}$Be($1/2^-$), and $^{11}$Be($3/2^-_2$) states
are described by $\pi_{3/2}^2\sigma_{1/2}$,  $\pi_{3/2}^2\pi_{1/2}^1$  and $\pi_{3/2}^1\sigma_{1/2}^2$ configurations,
while $^{12}$Be($0^+_1$), $^{12}$Be($0^+_2$), and $^{12}$Be($1^-_1$) states
correspond to $\pi_{3/2}^2\sigma_{1/2}^2$,  $\pi_{3/2}^2\pi_{1/2}^2$  and $\pi_{3/2}^2\pi_{1/2}^1\sigma_{1/2}^1$, respectively. 
These assignments have been suggested by molecular orbital models \cite{Oertzen-rev,ITAGAKI},
cluster models \cite{Arai01,Arai:2004yf,Ito:2003px,Ito:2005yy,Ito:2008zza}, and 
also the AMD model \cite{Dote:1997zz,ENYObe10,ENYObe11,ENYObe12}.
For $^{13}$Be, the molecular orbital configurations, $\pi_{3/2}^2\pi_{1/2}^1\sigma_{1/2}^2$ and  $\pi_{3/2}^2\sigma_{1/2}^2\pi^{*1}_{3/2}$, are assigned to 
$^{13}$Be($1/2^-$) and $^{13}$Be($3/2^+$), respectively, and 
$\pi_{3/2}^2\pi_{1/2}^2\sigma_{1/2}^1$ is temporarily assigned to $^{13}$Be($5/2^+$), as mentioned.
For $^{14}$Be, the ground state $^{14}$Be($0^+_1$) is considered to be 
a $\pi_{3/2}^2\pi_{1/2}^2\sigma_{1/2}^2$ configuration. T
An excited $0^+_2$ band was theoretically suggested by a VAP calculation\cite{ENYObe14} and a
$\pi_{3/2}^2\sigma_{1/2}^2\pi^{*2}_{3/2}$ configuration was assigned. 

\begin{figure}[th]
\epsfxsize=7 cm
\centerline{\epsffile{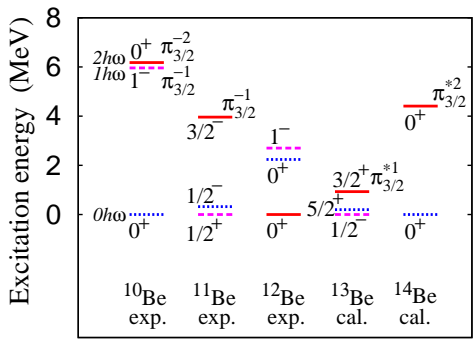}}
\caption{ (Color on-line)
Excitation energies of band-head states in $^{10-14}$Be. 
Experimental data are shown for $^{10-12}$Be, and theoretical values
for $^{14}$Be\cite{ENYObe14} are shown.
The data for $^{13}$Be are the present results obtained by the $^{12}$Be-$n$ model calculation.
Blue dotted, magenta dashed, and red solid lines indicate 
$0\hbar\omega$,  $1\hbar\omega$, and  $2\hbar\omega$ configuration states, respectively.
\label{fig:beiso-bh}
}
\end{figure}

Finally, we discuss energy spectra of Be isotopes in relation with molecular orbital configurations.
In Fig.~\ref{fig:beiso-bh}, experimental excitation energies are shown 
for $^{10-12}$Be, while theoretical values are shown for  $^{13}$Be and $^{14}$Be\cite{ENYObe14}.
In terms of spherical shell model levels, 
the inversion
between $0\hbar\omega$, $1\hbar\omega$, and $2\hbar\omega$ occurs in $^{11-13}$Be, and 
the energy spectra seem to be out of the normal ordering.
However, in terms of MO levels, the energy spectra of Be isotopes can be understood rather easily.
In MO levels, the neutron Fermi level exists at the
$\sigma_{1/2}$-$\pi_{1/2}$ shell in $^{10-14}$Be.  Excited configurations are characterized by 
$\pi_{3/2}$-orbital holes or $\pi^*_{3/2}$-orbital particles. 
Configurations without $\pi^*_{3/2}$ particles nor $\pi_{3/2}$ holes are normal in the MO levels 
and they degenerate in a low-energy region. For instance, the degeneracy of $^{11}$Be$(1/2^+)$ and 
$^{11}$Be$(1/2^-)$
can be understood because all these states have ``normal" configurations in MO levels.
Also the coexistence of $^{12}$Be$(0^+_1)$, $^{12}$Be$(0^+_2)$, and $^{12}$Be$(1^-)$ in the low-energy region
is to be expected because they have no excitation in MO configurations. 
In a similar way, the coexistence of $^{13}$Be$(1/2^-)$ and $^{13}$Be$(5/2^+)$ is not surprising 
as they have normal MO configurations.
Excited configuration states with  $\pi_{3/2}$ holes or $\pi^*_{3/2}$ particles have 
generally higher excitation energies than low-lying normal MO configuration states.
The numbers of $\pi_{3/2}$ holes and $\pi^*_{3/2}$ particles are noted in Fig.~\ref{fig:beiso-bh} as well as Table.~\ref{tab:beiso}. 
$^{10}$Be$(0^+_2)$, $^{10}$Be$(1^-)$, $^{10}$Be$(0^+_2)$, and $^{11}$Be$(3/2^-)$ have one or two $\pi_{3/2}$ holes,
and they exist in the excitation energy region around $E_x=4\sim 6$ MeV.
The excited state with a $\pi^*_{3/2}$ particle and that with two $\pi^*_{3/2}$ particles are suggested in 
$^{13}$Be$(3/2^+)$ and $^{14}$Be$(0^+_2)$, respectively.
For more detailed discussion of energy spectra, 
one should take into account two-body correlations and $\alpha$-$\alpha$ distance dependence of those MO levels.

\begin{table}[ht]
\label{tab:beiso}
\caption{Classification of band-head states in $^{10-14}$Be. Harmonic oscillator shell-model
configurations and molecular orbital configurations are listed. 
The shell model configurations $0\hbar\omega$, $1\hbar\omega$, and $2\hbar\omega$
 are based on neutrons excited from the $p$-shell to $sd$-shell.
For molecular orbital configurations, neutron configurations around a $2\alpha$ core are 
described. 
The numbers of 
$\pi_{3/2}$ holes and $\pi^*_{3/2}$ particles are also shown.  
The molecular orbital configuration for $^{13}$Be($5/2^+$) is a temporary assignment (See text).
}
\begin{center}
\begin{tabular}{c|c|ccc}
\hline
 & H.O. &  \multicolumn{3}{c}{M.O.} \\
 &  &   & $\pi_{3/2}$ & $\pi^*_{3/2}$ \\
  & excitation & config. & holes & particles \\
\hline
$^{10}$Be($0^+_1$) &  $0\hbar\omega$ & $\pi^2_{3/2}$ & 0 & 0 \\
$^{10}$Be($1^-$) &  $1\hbar\omega$ & $\pi_{3/2}\sigma_{1/2}$ & 1 & 0 \\
$^{10}$Be($0^+_2$) &  $2\hbar\omega$ & $\sigma^2_{1/2}$ & 2 & 0 \\
\hline
$^{11}$Be($1/2^+$) &  $1\hbar\omega$ & $\pi^2_{3/2}\sigma_{1/2}$ & 0 & 0 \\
$^{11}$Be($1/2^-$) &  $0\hbar\omega$ & $\pi^2_{3/2}\pi_{1/2}$ & 0 & 0 \\
$^{11}$Be($3/2^-$) &  $2\hbar\omega$ & $\pi_{3/2}\sigma^2_{1/2}$ & 1 & 0 \\
\hline
$^{12}$Be($0^+_1$) &  $2\hbar\omega$ & $\pi^2_{3/2}\sigma^2_{1/2}$ & 0 & 0 \\
$^{12}$Be($0^+_2$) &  $0\hbar\omega$ & $\pi^2_{3/2}\pi^2_{1/2}$ & 0 & 0 \\
$^{12}$Be($1^-$) &  $1\hbar\omega$ & $\pi^2_{3/2}\pi_{1/2}\sigma_{1/2}$ & 0 & 0 \\
\hline
$^{13}$Be($1/2^-$) &  $1\hbar\omega$ & $\pi^2_{3/2}\pi_{1/2}\sigma^2_{1/2}$ & 0 & 0 \\
$^{13}$Be($5/2^+$) &  $0\hbar\omega$ & $(\pi^2_{3/2}\pi^2_{1/2}\sigma_{1/2})$ & 0 & 0 \\
$^{13}$Be($3/2^+$) &  $2\hbar\omega$ & $\pi^2_{3/2}\sigma^2_{1/2}\pi^*_{3/2}$ & 0 & 1 \\
\hline
$^{14}$Be($0^+_1$) &  $0\hbar\omega$ & $\pi^2_{3/2}\pi^2_{1/2}\sigma^2_{1/2}$ & 0 & 0 \\
$^{14}$Be($0^+_2$) &  $2\hbar\omega$ & $\pi^2_{3/2}\sigma^2_{1/2}\pi^{*2}_{3/2}$ & 0 & 2 \\
\hline
\end{tabular}
\end{center}
\end{table}

\section{Summary}\label{sec:summary}

Structure of $^{13}$Be was investigated with VAP+AMD. 
In the AMD+VAP calculation using the set (1) interaction that reproduces the parity inversion of
$^{11}$Be, 
an unnatural parity $1/2^-$ state was suggested to be lower than $5/2^+$ state indicating that
the vanishing of $N=8$ magic number occurs in $^{13}$Be.
A low-lying $3/2^+$ state with a $2\hbar\omega$ configuration was also suggested. 
The present AMD+VAP calculation is a bound state approximation. 
To see effects of spatial extension of the last neutron wave function
on the energy spectra, we also calculated $^{13}$Be in the $^{12}$Be+$n$ model. 
The degeneracy of the $1/2^-$, $5/2^+$, and $3/2^+$ states was suggested also by the 
$^{12}$Be+$n$ model calculation.
It is necessary to treat asymptotic behaviors and out-going boundary more carefully
to discuss detailed energy spectra of resonances. 

In analysis of intrinsic structures of $^{13}$Be, 
large deformations with developed cluster structures are found in 
the intruder states such as  the $1/2^-$ and $3/2^+$ states. 
These deformed states are regarded as states composed by
one neutron on the deformed $^{12}$Be core, which corresponds to the intrinsic 
state of $^{12}$Be$(0^+_1)$.

The intrinsic structures of $^{13}$Be were also discussed 
in terms of molecular orbitals around a $2\alpha$ core.
The intruder states, $^{13}$Be($1/2^-$) and $^{13}$Be($3/2^+$) are described by 
the molecular configurations, 
$\pi_{3/2}^2\pi_{1/2}^1\sigma_{1/2}^2$ and  $\pi_{3/2}^2\sigma_{1/2}^2\pi^{*1}_{3/2}$,
respectively. 
The breaking of $N=8$ magicity in $^{13}$Be 
can be understood by molecular orbital levels
as well as $^{11}$Be and $^{12}$Be.

\end{document}